\documentclass[]{elsarticle}
\usepackage{lineno,hyperref}
\usepackage{epstopdf} %!!!!!!
\usepackage{color}
\usepackage{lscape}
\modulolinenumbers[5]
\biboptions{sort&compress}
\journal{Acta Materialia}
\usepackage{graphicx}% Include figure files
\RequirePackage{amsmath}
\begin{document}
\begin{frontmatter}
\title{Thermodynamic potentials of metallic alloys in the undercooled liquid  and  solid glassy states}

\author[mymainaddress]{A.S. Makarov}
\author[mymainaddress]{R.A. Konchakov}
\author[mymainaddress3]{N.P. Kobelev}
\author[mymainaddress]{V.A. Khonik}
\corref{mycorrespondingauthor}
\ead{v.a.khonik@yandex.ru}

\address[mymainaddress]{Department of General Physics, State Pedagogical University, Lenin Street 86, Voronezh 394043, Russia}
\address[mymainaddress3]{Institute of Solid State Physics, Russian Academy of Sciences, Chernogolovka, Moscow district 142432, Russia}

\begin{abstract}

We first present a comparative analysis of  temperature evolution of the excess thermodynamic potentials (state functions), the enthalpy $\Delta H$, entropy $\Delta S$ and Gibbs free energy $\Delta \Phi$, determined for \textit{i}) undercooled melts  using literature data and \textit{ii}) solid glassy state calculated on the basis of calorimetry measurements  using an approach proposed recently. Three metallic alloys  were taken as an example for data analysis. It is found that temperature dependences $\Delta H(T)$, $\Delta S(T)$ and $\Delta G(T)$ calculated with both approaches coincide in the supercooled liquid range (i.e. at temperatures $T_g<T<T_x$, where $T_g$ and $T_x$ are the glass transition and crystallization onset temperatures, respectively). However, the necessary conditions for this coincidence is the introduction of important changes to the above approach \textit{i}), which are related to the calculation of the melting entropy.   We also introduce and calculate a dimensionless order parameter $\xi$, which changes in the range $0<\xi<1$ and characterizes the evolution of the structural order from liquid-like to crystal-like  one. It is shown that the order parameter $\xi_{scl}$ calculated for the end of the supercooled liquid range (i.e. for a temperature just below $T_x$) correlates with the melt critical cooling rate $R_c$: the smaller the order parameter $\xi_{scl}$ (i.e. the closer the structure to that of the equilibrium liquid), the smaller $R_c$ is.      

\end{abstract}

\begin{keyword}
metallic glasses, calorimetry,  thermodynamic potentials, order parameter, critical cooling rate 
\end{keyword}
\end{frontmatter}

\section{Introduction}
The appearance of metallic glasses (MGs) produced by rapid melt solidification has sparked a new interest in the study of nucleation and crystal growth kinetics in undercooled melts \cite{Karthika2016,NaPNAS2020,PrabhuPhysicaB2022,KuballIntermetallics2024}. An important parameter in the theory of nucleation is the difference between the Gibbs free energy of a melt and that of a solid crystalline phase. This difference is  known as the excess Gibbs free energy of the liquid relative to the crystalline state \cite{Frank1952,Turnbull1949,Turnbull1950}. The fundamental thermodynamic relations allow us to calculate the excess Gibbs free energy of a liquid, $\Delta \Phi_{L-X}(T)$,  with respect to its crystalline state based on the known difference in specific heat capacities between the liquid and solid phases at constant pressure, $\Delta C_p^{L-X}(T) = C_p^L(T) - C_p^X(T)$ \cite{Chang2009}. Meanwhile, the temperature dependence of the specific heat difference  $\Delta C_{P}^{L-X}(T)$ over the entire range of the undercooled melt's existence is rarely available. At best, the specific heat difference $\Delta C_{p}^{L-X}(T)$ can be measured near the melting point $T_f$ as well as near the glass transition temperature $T_g$. 

Therefore, models have been developed to estimate the excess Gibbs free energy $\Delta\Phi_{L-X}(T)$, which include  polynomial approximations of $\Delta C_{p}^{L-X}(T)$ and experimentally determinable parameters such as the melting point $T_{f}$, enthalpy of melting $\Delta H_{f}$ and entropy of melting $\Delta S_{f}$. Currently, the most popular approach uses experimental data on the specific heat of the melt $C_{p}^{L}(T)$ and crystal $C_{p}^{X}(T)$ near $T_f$ and $T_g$, which approximated by the polynomials \cite{HoffmanJChemPhys1958}

\begin{equation}
C_{p}^{L}(T)=3R+aT+bT^{-2}, \label{C_{PL}}
\end{equation}

\begin{equation}
C_{p}^{X}(T)=3R+cT+dT^{2}, \label{C_{PX}}
\end{equation}
where  $R$ is the universal gas constant, $a$ and $b$ are the fitting parameters for the melt ($L$), $c$ and $d$ are the fitting parameters for the crystal ($X$). Polynomials (\ref{C_{PL}}) and (\ref{C_{PX}}) allow calculating temperature dependence of the difference $\Delta C_{P}^{L-X}(T)$  between the specific heat of the supercooled melt and that of the crystal over the entire range of the existence of the supercooled melt \cite{Busch1995,Busch1998,Busch2000,Legg2007,Jiang2008}. In turn, the quantity $\Delta C_{P}^{L-X}(T)$  allows determination of the excess enthalpy $\Delta H_{L-X}(T)$, excess entropy $\Delta S_{L-X}(T)$ and excess Gibbs free energy $\Delta\Phi_{L-X}(T)$ of the undercooled melt with respect to the crystal using the relations \cite{Thompson1979}:
     
\begin{equation}
\Delta H_{L-X}(T)=\Delta H_{f}-\int_{T}^{T_{f}} \Delta C_{p}^{L-X}(T)dT, \label{H_{L-X}}
\end{equation}

\begin{equation}
\Delta S_{L-X}(T)=\Delta S_{f}-\int_{T}^{T_{f}} \frac{\Delta C_{p}^{L-X}(T)}{T}dT, \label{S_{L-X}}
\end{equation}

\begin{equation}
\Delta \Phi_{L-X}(T)=\Delta H_{L-X}(T)-T \Delta S_{L-X}(T), \label{Phi_{L-X}}
\end{equation}
where all quantities are described above.

On the other hand, equations defining the excess thermodynamic potentials of MGs both in the \textit{solid} (i.e. below $T_g$) and supercooled liquid (i.e. above $T_g$) states  with respect to the crystalline counterpart are proposed in Refs \cite{MakarovJPCM2021,MakarovJETPLett2022,MakarovJETPLett2024}. The relations for excess enthalpy $\Delta H_{G-X}(T)$, excess entropy $\Delta S_{G-X}(T)$, and excess Gibbs free energy $\Delta\Phi_{G-X}(T)$ of glass with respect to the crystal  have the form \cite{MakarovJPCM2021,MakarovJETPLett2022,MakarovJETPLett2024}: 

\begin{equation}
\Delta H_{G-X}(T)=\frac{1}{\dot{T}}\int_{T}^{T_{cr}} \Delta W_{G-X}(T)dT, \label{H_{G-X}}
\end{equation}

\begin{equation}
\Delta S_{G-X}(T)=\frac{1}{\dot{T}}\int_{T}^{T_{cr}} \frac{\Delta W_{G-X}(T)}{T}dT, \label{S_{G-X}}
\end{equation}

\begin{equation}
\Delta \Phi_{G-X}(T)=\int_{T}^{T_{cr}} \Delta S_{G-X}(T)dT, \label{Phi_{G-X}}
\end{equation}
where $\dot{T}$ is the heating rate, $T_{cr}$ is the temperature of the complete crystallization, $\Delta W_{G-X}(T)$ constitutes the difference between the heat flow coming from glass $W_{G}(T)$ and its crystalline counterpart $W_{X}(T)$ and hereafter termed as the differential heat flow, i.e. $\Delta W_{G-X}(T)=W_{G}(T)-W_{X}(T)$. These equations allow calculating temperature dependences $\Delta H_{G-X}(T)$, $\Delta S_{G-X}(T)$ and $\Delta \Phi_{G-X}(T)$ from a low temperature (e.g. from room temperature) up to $T_{cr}$. It should be emphasized that if current temperature $T=T_{cr}$, then the integrals (\ref{H_{G-X}})--(\ref{Phi_{G-X}}) become zero, and therefore, the quantities   $\Delta H_{G-X}$, $\Delta S_{G-X}$ and $\Delta \Phi_{G-X}(T)$ describe solely the excess enthalpy, entropy and Gibbs free energy of the glass compared to its crystalline counterpart. 

While the analysis based on Eqs (\ref{H_{L-X}})--(\ref{Phi_{L-X}}) has been used for a long time  \cite{Thompson1979}, relations (\ref{H_{G-X}})--(\ref{Phi_{G-X}}) constitute a novel approach, which provides new interesting results \cite{MakarovJPCM2021,MakarovJETPLett2022,MakarovJETPLett2024,KonchakovJETPLett2024,MakarovScrMater2024,MakarovIntermet2024,AfoninJALCOM2024b}.
It is, therefore, important to check whether these two approaches give consistent data in the common temperature range (i.e. at temperatures from $T_g$ up to the crystallization onset temperature $T_x$). This is the first goal of the present investigation. Besides that, we perform a comparative analysis of these two methods in the explanation of MGs' glass-forming ability and revealed their advantages and shortcomings. 

\section{Experimental}

The experimental part of the present investigation was performed on bulk glassy (at.\%) Pd$_{40}$Ni$_{40}$P$_{20}$, Pt$_{42.5}$Cu$_{27}$Ni$_{9.5}$P$_{21}$ and high-entropy Zr$_{35}$Hf$_{13}$Al$_{11}$Ag$_{8}$Ni$_{8}$Cu$_{25}$ (the mixing entropy $\Delta S_{mix}/R$=1.63). The first alloy was prepared by direct melting of Pd (purity 99.95\%) and Ni$_2$P (purity 99.5\%) using alumina crucible in a pure Ar atmosphere and next treated in B$_2$O$_3$ flux. The second alloy was produced by direct remelting of the constituent elements (at least 99.9\% pure) in evacuated quartz vial using a two-temperature method. High-entropy Zr$_{35}$Hf$_{13}$Al$_{11}$Ag$_{8}$Ni$_{8}$Cu$_{25}$ alloy was prepared by direct remelting of the elements (purity not worse than 99.9\%) in  an induction furnace in a pure Ar atmosphere. All glasses were prepared by  melt jet quenching.  The non-crystallinity of samples was verified by X-rays. For all compositions, the data on the specific heat of the melt ($C_{p}^{L}$) and crystal ($C_{p}^{X}$), melting point ($T_f$) and glass transition temperature ($T_g$) are available in the literature \cite{Wilde1994,Lu1999,Neuber2021,Ohashi2022}. 

Excess thermodynamic potentials (Eqs (\ref{H_{G-X}})--(\ref{Phi_{G-X}}))  were determined for the relaxed (preannealed) state of all MGs using differential scanning calorimetry (DSC) data derived by a Hitachi DSC 7020 instrument operating in high-purity (99.999 \%) N$_{2}$ atmosphere at a rate of 3 K/min. Every glass was tested according to the following protocol: \textit{i}) an initial sample was heated up to the temperature of the complete crystallization $T_{cr}$ with an empty reference DSC cell; this crystallized sample was next moved to the reference DSC cell; \textit{ii}) a new sample in the initial state was heated into the supercooled liquid state (i.e. into the range $T_g\leq T < T_x$, where  $T_{x}$ is the crystallization onset temperature) (run 1) and cooled back to room temperature at about same rate of  3 K/min; \textit{iii}) the same (relaxed) sample was tested up to $T_{cr}$ (run 2) and, finally, \textit{iv}) the same (crystallized) specimen is again heated up to $T_{cr}$ (run 3). This protocol allows performing DSC measurements with the reference cell containing fully crystallized sample of approximately the same mass (50--70 mg) so that the measured heat flow  during run 3 constitutes the difference between the heat flow coming from the relaxed glass $W_{G}(T)$ and its crystalline counterpart $W_{X}(T)$, i.e. the differential heat flow $\Delta W_{G-X}(T)=W_{G}(T)-W_{X}(T)$. The DSC instrument was calibrated using the melting points and enthalpies of 99.99 \% pure In, Sn and Pb. The uncertainty in the enthalpy changes was estimated to be about 4\%. The nitrogen flow rate was 60 ml/min. Under the above conditions, no visible oxidation of samples was detected.

\section{Results and discussion}

\subsection{Thermodynamic potentials of undercooled  melts}

\begin{figure}[t]
\center{\includegraphics[scale=1.1]{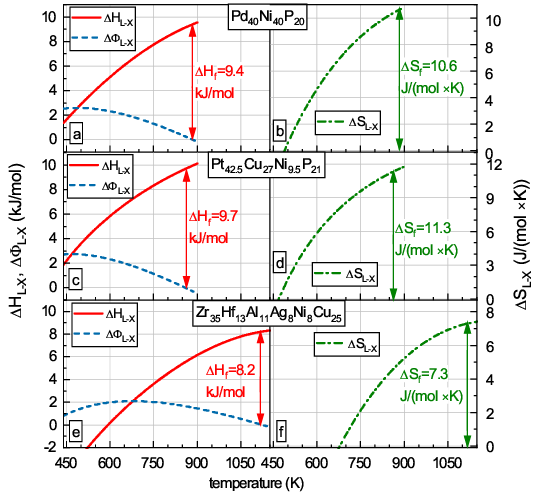}}% Here is how to import EPS art
\caption[*]{\label{Fig1.eps} Temperature dependences of the excess enthalpy $\Delta H_{L-X}$, excess entropy $\Delta S_{L-X}$ and excess Gibbs free energy $\Delta\Phi_{L-X}$  of Pd$_{40}$Ni$_{40}$P$_{20}$  (a,b) \cite{Wilde1994,Lu1999}), Pt$_{42.5}$Cu$_{27}$Ni$_{9.5}$P$_{21}$ (c,d)  \cite{Neuber2021} and Zr$_{35}$Hf$_{13}$Al$_{11}$Ag$_{8}$Ni$_{8}$Cu$_{25}$ (e,f)  \cite{Ohashi2022} undercooled melts with respect to the crystalline state obtained with equations (\ref{H_{L-X}})--(\ref{Phi_{L-X}}). The enthalpy $\Delta H_{f}$ (Eq. \ref{DeltaHf}) and entropy $\Delta S_{f}$ (Eq.(\ref{DeltaSf})) at  liquidus temperatures $T_{L}$ are shown by the red and green arrows, respectively.}
\end{figure}

Figure \ref{Fig1.eps} shows  literature data on  temperature dependences of the excess enthalpy $\Delta H_{L-X}(T)$, excess entropy $\Delta S_{L-X}(T)$ and excess Gibbs free energy $\Delta\Phi_{L-X}(T)$ for undercooled melts of Pd$_{40}$Ni$_{40}$P$_{20}$ \cite{Wilde1994,Lu1999} (a,b), Pt$_{42.5}$Cu$_{27}$Ni$_{9.5}$P$_{21}$ \cite{Neuber2021} (c,d) and Zr$_{35}$Hf$_{13}$Al$_{11}$Ag$_{8}$Ni$_{8}$Cu$_{25}$ \cite{Ohashi2022} (e,f) with respect to the crystalline state obtained with equations (\ref{H_{L-X}})--(\ref{Phi_{L-X}}).

As an example, let us consider the data on Pt$_{42.5}$Cu$_{27}$Ni$_{9.5}$P$_{21}$ alloy shown in
Fig.\ref{Fig1.eps} (c,d). The specific heat of the glassy, liquid and crystalline states of this alloy was studied in Ref.\cite{Neuber2021}. Temperature dependences of the  heat capacities were approximated by polynomials (\ref{C_{PL}}) and (\ref{C_{PX}}) with $a=11.5852\times{10^{-3}}$ Jmol$^{-1}$K$^{-2}$, $b=5.1317\times{10^{6}}$ Jmol$^{-1}$K, $c=-7.2702\times{10^{-3}}$ Jmol$^{-1}$K$^{-2}$ and $d=16.5269\times{10^{-6}}$ Jmol$^{-1}$K$^{-3}$ \cite{Neuber2021}. After that, temperature dependence of the difference in the specific heats of the supercooled melt and crystal $\Delta C_{p}^{L-X}(T)$ was determined and further used  to calculate the excess thermodynamic potentials of the supercooled melt using equations (\ref{H_{L-X}})--(\ref{Phi_{L-X}}). 

The melting temperature $T_{f}$ and melting enthalpy $\Delta H_{f}$ were determined by DSC method at a heating rate of 20 K/min. The melting temperature was accepted as the liquidus temperature $T_{L}$. The melting enthalpy   $\Delta H_{f}$ was determined from the heat flow $W(t,T)$ using the equation 
\begin{equation}
\Delta H_{f}=\int_{T_{S}}^{T_{L}} \frac{W(t,T)}{\dot{T}}dT, \label{DeltaHf}
\end{equation}
where $T_{S}$=799 K and $T_{L}$=863 K are the solidus and liquidus temperatures, respectively. The entropy of melting $\Delta S_{f}$ was defined as the ratio of the enthalpy of melting $\Delta H_{f}$ to the liquidus temperature,
\begin{equation}
\Delta S_{f}=\frac{\Delta H_{f}}{T_{L}}. \label{DeltaSf}
\end{equation}

Next, using Eqs (\ref{DeltaHf}) and (\ref{DeltaSf}), the quantities  $\Delta H_{f}$=9.7 kJmol$^{-1}$ and $\Delta S_{f}$=11.3 Jmol$^{-1}$K$^{-1}$ were obtained. The enthalpy of melting $\Delta H_{f}$ and entropy of melting $\Delta S_{f}$ at the liquidus temperatures $T_{L}$ are indicated in Fig.\ref{Fig1.eps}(c,d) by the red and green arrows, respectively. It is seen that the excess enthalpy and entropy of Pt$_{42.5}$Cu$_{27}$Ni$_{9.5}$P$_{21}$  melt decrease as the undercooling increases.
The excess Gibbs free energy $\Delta\Phi_{L-X}(T)$ is plotted in Fig.\ref{Fig1.eps}(c). 

Similar data on the thermodynamic potentials for other two alloys are given in panels (a,b) and (e,f)  of Fig.\ref{Fig1.eps}. The main conclusion from the data exemplified by Fig.\ref{Fig1.eps}  is based on $\Delta \Phi_{L-X}$-values taken at temperatures  $(0.4-0.6)\times T_{L}$, where $\Delta \Phi_{L-X}$ reaches its maximum. It is believed that the small excess Gibbs free energy  near these maxima reflect the small driving force for melt crystallization and, therefore, is crucial for understanding the high glass-forming ability of these melts
 \cite{Busch1998,Busch2000,Jiang2008,Neuber2021,Ohashi2022}. We will return to this conclusion below and show that it is not supported by the experimental data analyzed in this work.

\subsection{Enthalpies of glasses and undercooled melts}

\begin{figure}[t]
\center{\includegraphics[scale=1.1]{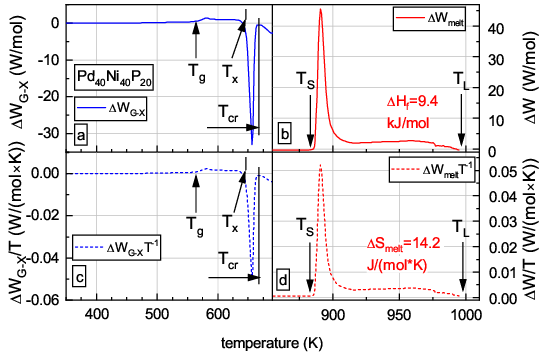}}% Here is how to import EPS art
\caption[*]{\label{Fig2.eps} (a) Differential heat flow $\Delta W_{G-X}(T)$ of glassy Pd$_{40}$Ni$_{40}$P$_{20}$ up to the temperature of the complete crystallization $T_{cr}$. (b) DSC thermogram $\Delta W(T)$ of the same alloy after crystallization up to the liquidus temperature $T_{L}$.
(c) The same data as in panel (a)  but divided by the absolute temperature, i.e. $\Delta W_{G-X}(T)/T$.  (d) The same data as in panel (b)  but divided by the absolute temperature, i.e. $\Delta W(T)/T$. The characteristic temperatures $T_g$ (glass transition), $T_x$ (crystallization onset), $T_{cr}$, $T_{S}$ (solidus) and $T_{L}$ are indicated.}
\end{figure}

Let us now  consider the enthalpy of the same alloys in the solid glassy state. Fig.\ref{Fig2.eps}(a) shows  temperature dependence of the differential heat flow $\Delta W_{G-X}(T)$ of relaxed  Pd$_{40}$Ni$_{40}$P$_{20}$ glass derived in the present investigation. The relaxation was performed by heating of initial  sample up to 610 K (which corresponds to the supercooled liquid state) and subsequent controlled cooling to room temperature. 

\begin{figure}[h]
\center{\includegraphics[scale=1.1]{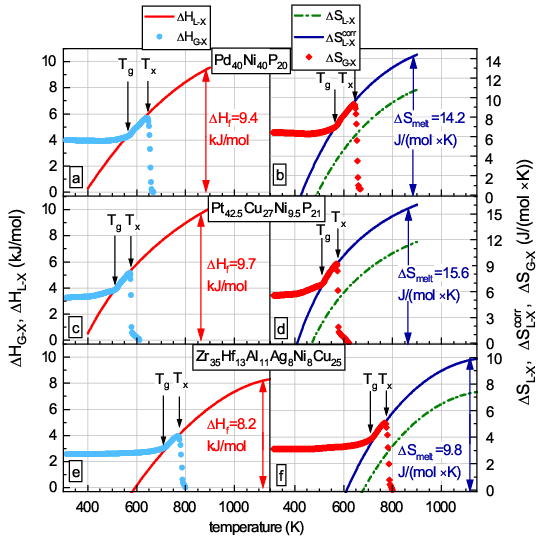}}% Here is how to import EPS art
\caption[*]{\label{Fig3.eps} Temperature dependences of the excess enthalpies and excess entropies for the undercooled melts (labelled $L-X$) and glasses (labelled $G-X$):  \textit{i}) the excess enthalpy $\Delta H_{L-X}(T)$, excess  entropy $\Delta S_{L-X}(T)$ and corrected excess entropy $\Delta S_{L-X}^{corr}(T)$ calculated using Eqs (\ref{H_{L-X}}), (\ref{S_{L-X}}) and (\ref{S_{L-X}^{corr}}), respectively;  \textit{ii}) the excess enthalpy $\Delta H_{G-X}(T)$ and excess entropy $\Delta S_{G-X}(T)$ determined with Eqs (\ref{H_{G-X}}) and (\ref{S_{G-X}}), respectively. 

Panels (a,b), (c,d) and (e,f) correspond to the alloys Pd$_{40}$Ni$_{40}$P$_{20}$, Pt$_{42.5}$Cu$_{27}$Ni$_{9.5}$P$_{21}$ and Zr$_{35}$Hf$_{13}$Al$_{11}$Ag$_{8}$Ni$_{8}$Cu$_{25}$, respectively.  The glass transition temperatures $T_{g}$ and crystallization onset temperatures $T_{x}$ are shown. The  melting enthalpy $\Delta H_{f}$ (Eq.\ref{DeltaHf}) and melting entropy $\Delta S_{melt}$ (Eq.(\ref{DeltaSmelt})) are indicated.}
\end{figure}

The data in Fig.\ref{Fig2.eps}(a) were used to calculate the excess enthalpy $\Delta H_{G-X}(T)$ with Eq.(\ref{H_{G-X}}), which is shown in Fig.\ref{Fig3.eps}(a). It is seen that $\Delta H_{G-X}(T)$ (light blue symbols)  remains approximately constant up to $T_{g}$, rapidly increases in the supercooled liquid state (i.e. at  $T_g\leq T\leq T_x$) and finally sharply fulls down to zero due to full crystallization at $T_{cr}=668$ K (indicated by the arrow in Fig. \ref{Fig2.eps}(a)). Red solid curve in Fig.\ref{Fig3.eps}(a) gives the excess melt enthalpy $\Delta H_{L-X}(T)$ for the same alloy system replotted from Refs \cite{Wilde1994,Lu1999}. It is seen that the glass excess enthalpy $\Delta H_{G-X}(T)$ and melt excess enthalpy $\Delta H_{L-X}(T)$ fully coincide in their common temperature range $T_{g}\leq T \leq T_{x}$. This means that these quantities agree with each other.  

Panels  (c) and (e) in Fig.\ref{Fig3.eps} show analogous data set for the alloys Pt$_{42.5}$Cu$_{27}$Ni$_{9.5}$P$_{21}$ and Zr$_{35}$Hf$_{13}$Al$_{11}$Ag$_{8}$Ni$_{8}$Cu$_{25}$, respectively. The excess entropy $\Delta H_{G-X}(T)$ for the glassy state (light blue curves) was calculated using Eq.(\ref{H_{G-X}}) on the basis of DSC data derived in the present investigation as described above. The excess entropy  $\Delta H_{L-X}(T)$ (red curves) for Pt$_{42.5}$Cu$_{27}$Ni$_{9.5}$P$_{21}$  and Zr$_{35}$Hf$_{13}$Al$_{11}$Ag$_{8}$Ni$_{8}$Cu$_{25}$ undercooled melts was taken from Refs \cite{Neuber2021} and \cite{Ohashi2022}, respectively. It is seen that the enthalpies $\Delta H_{G-X}$ and  $\Delta H_{L-X}$ coincide in their common range $T_{g}\leq T \leq T_{x}$, just in the above case of Pd$_{40}$Ni$_{40}$P$_{20}$ alloy.

\subsection{Entropies of glasses and undercooled melts}

To calculate the excess entropy of glass $\Delta S_{G-X}$ with Eq.(\ref{S_{G-X}}), one needs the data on the differential heat flow divided by temperature, i.e. $\Delta W_{G-X}/T$, which is shown in   Fig.\ref{Fig2.eps}(c) for glassy Pd$_{40}$Ni$_{40}$P$_{20}$. The result of this calculation is given  in Fig.\ref{Fig3.eps}(b) by red symbols. In a qualitative sense, temperature change of the entropy  $\Delta S_{G-X}(T)$ follows the same pattern as the excess enthalpy described above (see Fig.\ref{Fig3.eps}(a), light blue symbols). For a comparison, we also plotted the melt excess entropy $\Delta S_{L-X}(T)$ derived in Refs.\cite{Wilde1994,Lu1999} for the same alloy system, see  dash-dotted green curve in Fig.\ref{Fig3.eps}(b). It is seen that the  melt excess entropy $\Delta S_{L-X}(T)$ is far from being equal to  the glass excess entropy $\Delta S_{G-X}(T)$ in the supercooled liquid region (i.e. in the range $T_g\leq T\leq T_x)$). 

The reason of this discrepancy seems to be quite clear. Upon calculation of the melt excess  enthalpy $\Delta H_{L-X}(T)$ with Eq.(\ref{H_{L-X}}), the authors \cite{Wilde1994,Lu1999} use the the melting enthalpy $\Delta H_{f}$, which was determined by integrating the heat flow data $W(t,T)$  according to  Eq.(\ref{DeltaHf}), and got a melting enthalpy of 9.4 kJ/mol.  This fully agrees with our experiment. Indeed, Fig.\ref{Fig2.eps}(b) shows the heat flow of  crystalline alloy Pd$_{40}$Ni$_{40}$P$_{20}$ in the temperature range 850 K $\leq T\leq 1000$ K. Then, the application of Eq.(\ref{DeltaHf}) yields the same melting enthalpy $\Delta H_{f}=9.4$ kJ/mol. Accordingly, the excess enthalpies of melt $\Delta H_{L-X}(T)$ and glass $\Delta H_{G-X}(T)$ coincide in the supercooled liquid state, as mentioned above. 

However, the situation with the melting entropies is different. To calculate the melting entropy $\Delta S_{f}$, the authors \cite{Wilde1994,Lu1999} used the ratio of the melting enthalpy to the solidus temperature given by Eq.(\ref{DeltaSf}) and arrive to a melting entropy $\Delta S_{f}$ = 10.6 Jmol$^{-1}$K$^{-1}$ as indicated in  Fig.\ref{Fig1.eps}(b). Next, using Eq.(\ref{S_{L-X}}) they determined temperature dependence of the melt excess entropy $\Delta S_{L-X}(T)$, which is shown by green dash-dotted line in Fig.\ref{Fig3.eps}(a).  It is seen that in the temperature range of interest ($T_g \leq T\leq T_x$) this entropy lies far below the excess entropy of glass $\Delta S_{G-X}(T)$ calculated with Eq.(\ref{S_{G-X}}) and shown by the red symbols in Fig.\ref{Fig3.eps}(b). We believe that this inconsistency is due to the incorrect use of Eq.(\ref{DeltaSf}) to calculate the melting entropy: according to general thermodynamics \cite{Landau}, the differential of the entropy $dS=\delta Q/T$ ($\delta Q$ is the elementary heat) and, therefore, the macroscopic entropy change must be calculated as an integral, $\Delta S = \int \delta Q/T$.  Consequently, the melting entropy should be accepted as \cite{MakarovScrMater2024}

\begin{equation}
\Delta S_{melt}=\frac{1}{\dot{T}}\int_{T_S}^{T_L}\frac{\Delta W(T)}{T}dT. \label{DeltaSmelt}
\end{equation}
Figure \ref{Fig2.eps}(d) shows $\Delta W(T)/T$-dependence  for crystalline Pd$_{40}$Ni$_{40}$P$_{20}$ in the range from 850 K to 1000 K. Using this dependence together with Eq.(\ref{DeltaSmelt}), one can find the melting  entropy $\Delta S_{melt}$ = 14.2 Jmol$^{-1}$K$^{-1}$.  

Now let us rewrite formula (\ref{S_{L-X}}) and define the corrected entropy difference between the undercooled liquid and crystalline phases as 

\begin{equation}
\Delta S_{L-X}^{corr}(T)=\Delta S_{melt}-\int_{T}^{T_{f}} \frac{\Delta C_{P}^{L-X}(T)}{T}dT, \label{S_{L-X}^{corr}}
\end{equation}
where $\Delta S_{melt}$ (determined by Eq.(\ref{DeltaSmelt})) is used instead of $\Delta S_{f}$ in Eq.(\ref{S_{L-X}}).  Next, using this equation together with  $\Delta C_{P}^{L-X}(T)$-data  from Refs \cite{Wilde1994,Lu1999}, one can calculate the corrected temperature dependence of melt excess entropy $\Delta S_{L-X}^{corr}(T)$. The $\Delta S_{L-X}^{corr}(T)$-dependence thus determined is shown by the dark blue curve  in Fig.\ref{Fig3.eps}(b). It is seen that the corrected melt excess entropy $\Delta S_{L-X}^{corr}(T)$ completely coincides with the glass  excess entropy $\Delta S_{G-X}(T)$ (red symbols) in the common temperature range $T_{g}\leq T \leq T_{x}$.

The calculations of glass excess entropies  $\Delta S_{G-X}(T)$ using  Eq.(\ref{S_{G-X}}) for MGs  Pt$_{42.5}$Cu$_{27}$Ni$_{9.5}$P$_{21}$ and  Zr$_{35}$Hf$_{13}$Al$_{11}$Ag$_{8}$Ni$_{8}$Cu$_{25}$ are shown in panels (d) and (f) in Fig.\ref{Fig3.eps}, respectively. These $\Delta S_{G-X}(T)$-dependencies in general are pretty much similar to the case of glassy  Pd$_{40}$Ni$_{40}$P$_{20}$  (see Fig.\ref{Fig3.eps}(b)). 

Panels (d) and (f) in this Figure also show the data on melt excess entropies  $\Delta S_{L-X}(T)$ for  Pt$_{42.5}$Cu$_{27}$Ni$_{9.5}$P$_{21}$ and  Zr$_{35}$Hf$_{13}$Al$_{11}$Ag$_{8}$Ni$_{8}$Cu$_{25}$ (dash-dotted green curves) derived in Refs. \cite{Neuber2021} and \cite{Ohashi2022}, respectively. Again, it is seen that excess entropies of melts $\Delta S_{L-X}(T)$ lie far below the  excess entropies $\Delta S_{G-X}(T)$ for corresponding glasses in the supercooled liquid region. The reason for this discrepancy seems to be the same as above: the melting entropies were erroneously determined as the ratios of the melting enthalpies to liquidus temperatures \cite{Neuber2021,Ohashi2022}, i.e. as $\Delta S_{f}= \Delta H_{f}/T_{L}$ (indicated by green arrows in Fig.\ref{Fig1.eps}(d,f)). 

Regretfully, DSC thermograms for Pt$_{42.5}$Cu$_{27}$Ni$_{9.5}$P$_{21}$ and Zr$_{35}$Hf$_{13}$Al$_{11}$Ag$_{8}$Ni$_{8}$Cu$_{25}$ alloys covering the melting region are not available. In the view of this, we corrected the melting entropy data in the following way. One can notice that the melting entropies $\Delta S_{f}$ and $\Delta S_{melt}$ for Pd$_{40}$Ni$_{40}$P$_{20}$ alloy  differ by 35 \%. Asssuming the same difference for Pt$_{42.5}$Cu$_{27}$Ni$_{9.5}$P$_{21}$ and Zr$_{35}$Hf$_{13}$Al$_{11}$Ag$_{8}$Ni$_{8}$Cu$_{25}$, one can accept the melting entropy for these alloys to be $S_{melt}$ = 15.6 and 9.8 Jmol$^{-1}$K$^{-1}$, respectively (see dark blue arrows in Fig.\ref{Fig3.eps}(d,f)). These $\Delta S_{melt}$-values together with $\Delta C_{P}^{L-X}(T)$-data from Refs \cite{Neuber2021,Ohashi2022} were further used to recalculate the melt excess entropy using Eq.(\ref{S_{L-X}^{corr}}). Figure \ref{Fig3.eps} (d,f) shows temperature $\Delta S_{L-X}(T)$-dependences thus determined for the last two allows (see dark blue curves). It is seen that the excess entropies of melts and corresponding glasses coincide in the supercooled liquid state.

Overall, one can conclude that after appropriate corrections the melt and glass entropies derived using different approaches  agree with each other in the common temperature range.

\subsection{Gibbs free energy of undercooled melts and glasses}

\begin{figure}[t]
\center{\includegraphics[scale=1.1]{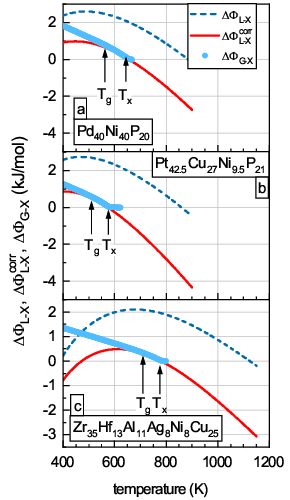}}% Here is how to import EPS art
\caption[*]{\label{Fig4.eps} Temperature dependences of the excess Gibbs free energy $\Delta\Phi$ of undercooled melts (labelled $L-X$) and metallic glasses (labelled $G-X$) calculated by Eqs  (\ref{Phi_{L-X}}), (\ref{Phi(L-X)(corr)}) and (\ref{Phi_{G-X}}) for the quantities $\Delta \Phi_{L-X}$, $\Delta\Phi_{L-X}^{corr}$ and $\Delta\Phi_{G-X}$, respectively. 
The data on $\Delta \Phi_{L-X}$ for Pd$_{40}$Ni$_{40}$P$_{20}$ (a), Pt$_{42.5}$Cu$_{27}$Ni$_{9.5}$P$_{21}$ (b)  and Zr$_{35}$Hf$_{13}$Al$_{11}$Ag$_{8}$Ni$_{8}$Cu$_{25}$ (c) alloys  are taken from  Refs \cite{Wilde1994,Lu1999}, \cite{Neuber2021} and \cite{Ohashi2022}, respectively. 
The glass transition temperatures $T_{g}$ and  crystallization onset temperatures $T_{x}$ are indicated.}
\end{figure}

Let us now consider the excess Gibbs free energy $\Delta\Phi$ of undercooled melts and glasses. Figure \ref{Fig4.eps} shows  literature data on temperature dependences of the excess  Gibbs free energy $\Delta\Phi_{L-X}(T)$ of Pd$_{40}$Ni$_{40}$P$_{20}$ \cite{Wilde1994,Lu1999}, Pt$_{42.5}$Cu$_{27}$Ni$_{9.5}$P$_{21}$ \cite{Neuber2021} and Zr$_{35}$Hf$_{13}$Al$_{11}$Ag$_{8}$Ni$_{8}$Cu$_{25}$ \cite{Ohashi2022} melts (dark blue dashed curves). This Figure also gives the Gibbs free energy  $\Delta\Phi_{G-X}(T)$ of same glasses calculated using Eq.(\ref{Phi_{G-X}}) (light blue curves). It is seen that the excess Gibbs free energy of undercooled melts $\Delta \Phi_{L-X}$  decreases with temperature at $T>T_{g}$ and becomes zero at the melting point. Meanwhile, the  excess Gibbs free energy of glasses $\Delta \Phi_{G-X}$ also decreases with temperature without any  features but goes to zero at $T=T_{cr}$. Thus, the excess Gibbs free energy $\Delta \Phi_{L-X}$
of undercooled melts is far different from the excess Gibbs free energy $\Delta \Phi_{G-X}$ of glasses. 

However, recalling incorrect determination of the melting entropy $\Delta S_{L-X}$  described above, one should replace it by the corrected entropy $\Delta S_{L-X}^{corr}$ in Eq.(\ref{Phi_{L-X}}), as argued in the previous Section. Then, instead of Eq.(\ref{Phi_{L-X}}) one should write down an expression for the corrected excess Gibbs free energy of undercooled melt as

\begin{equation}
\Delta \Phi_{L-X}^{corr}(T)=\Delta H_{L-X}(T)-T \Delta S_{L-X}^{corr}(T), \label{Phi(L-X)(corr)}
\end{equation}
where all the quantities are described above. 

Then, taking $\Delta H_{L-X}(T)$ from Refs  \cite{Wilde1994,Lu1999,Neuber2021,Ohashi2022} together with corrected $\Delta S_{L-X}^{corr}(T)$ and using Eq.(\ref{Phi(L-X)(corr)}), one can determine  corrected temperature dependence of the excess Gibbs free energy $\Delta \Phi_{L-X}^{corr}(T)$ of undercooled melts. The corresponding results are given by the red curves in Fig.\ref{Fig4.eps}. It is seen that corrected $\Delta \Phi_{L-X}^{corr}(T)$ for the alloys under consideration coincides with the excess Gibbs  free energy of corresponding glasses  $\Delta \Phi_{G-X}(T)$ in the supercooled liquid region, i.e. at temperatures $T_{g}\leq T \leq T_{x}$. \footnote{It is to be noted that, generally speaking,  Eqs (\ref{Phi_{L-X}}) and (\ref{Phi(L-X)(corr)}) cannot be applied for a description of an undercooled melt. Indeed, it is easy to show that  thermodynamic relationship 
\begin{equation}
\Delta \Phi=\Delta H - T\Delta S \label{*}
\end{equation}
 is valid only for isothermal conditions. Meanwhile, Eqs (\ref{Phi_{L-X}}) and (\ref{Phi(L-X)(corr)}) assume a varying temperature. For the latter case, by the definition of the Gibbs free energy \cite{Landau}, one should write down that $d\Delta \Phi=-\Delta SdT$ and, therefore, 
\begin{equation}
 \Delta \Phi=-\int \Delta S dT, \label{**}
 \end{equation} 
 which is similar to the relationship (\ref{Phi_{G-X}})  used to calculate  glass excess Gibbs free energy \cite{MakarovJETPLett2024}. However, numerical calculations with Eqs (\ref{*}) and (\ref{**}) give quite close results. } Above the temperature of complete crystallization $T_{cr}$, the Gibbs free energy $\Delta \Phi_{L-X}^{corr}$ becomes negative. This indicates negative work on the transformation of an undercooled melt into the crystal in the range $T_{cr}\leq T \leq T_{L}$.  
 
\subsection{On the correlation of  undercooled melt Gibbs free energy and melt critical cooling rate}

\begin{figure}[t]
\center{\includegraphics[scale=1.1]{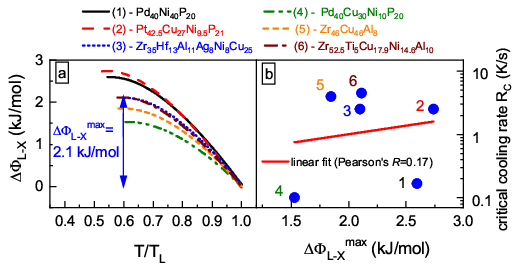}}% Here is how to import EPS art
\caption[*]{\label{Fig5.eps} (a) -- Temperature dependences of the excess Gibbs free energy $\Delta\Phi(T)_{L-X}$ of indicated undercooled melts as a function of the normalized temperature $T/T_L$. The maximal value on this dependence,  $\Delta\Phi_{L-X}^{max}$,  for Zr$_{35}$Hf$_{13}$Al$_{11}$Ag$_{8}$Ni$_{8}$Cu$_{25}$ is indicated. (b) -- The critical cooling rate $R_c$ of the same melts as a function of $\Delta\Phi(T)_{L-X}^{max}$. No clear correlation is seen.}
\end{figure}

As mentioned above, it is suggested in the literature that  small excess Gibbs free energy $\Delta \Phi_{L-X}$ at temperatures  $(0.4-0.6)\times T_{L}$, corresponding to a maximum of this dependence, reflect a  small driving force for melt crystallization and, therefore, should be indicative of good glass-forming ability (GFA) (e.g. Refs  \cite{Busch1998,Busch2000,Jiang2008,Neuber2021,Ohashi2022}). One can verify this suggestion for the alloys under consideration. 

Figure \ref{Fig5.eps}(a) shows  $\Delta \Phi_{L-X}$ as a function of the normalized temperature $T/T_{L}$. It is seen that, for instance, the maximum $\Delta\Phi_{L-X}^{max}$-value  for undercooled Zr$_{35}$Hf$_{13}$Al$_{11}$Ag$_{8}$Ni$_{8}$Cu$_{25}$ melt  is  2.1 kJmol$^{-1}$. Panel (b) in this Figure gives  the critical cooling rate $R_{c}$, which is a direct measure of the GFA, plotted as a function of $\Delta\Phi_{L-X}^{max}$. The $R_c$-data for  Zr$_{35}$Hf$_{13}$Al$_{11}$Ag$_{8}$Ni$_{8}$Cu$_{25}$, Pt$_{42.5}$Cu$_{27}$Ni$_{9.5}$P$_{21}$ and Pd$_{40}$Ni$_{40}$P$_{20}$  alloys are taken from Refs \cite{InoueMaterTransJIM1997,Gross2017,Ohashi2022}, respectively. The data on $\Delta\Phi_{L-X}^{max}$ and $R_c$ for Pd$_{40}$Cu$_{30}$Ni$_{10}$P$_{20}$ \cite{Lu1999,InoueMaterTransJIM1997}, Zr$_{46}$Cu$_{46}$Al$_{8}$ \cite{Jiang2008,XingJALCOM2004} and Zr$_{52.5}$Cu$_{17.9}$Ni$_{14.6}$Al$_{10}$Ti$_{5}$ \cite{MotykaMaterTrans2002,XingJALCOM2004} are taken from indicated references.

It is seen that \textit{i}) any clear correlation between $\Delta\Phi_{L-X}^{max}$ and $R_c$ is absent and \textit{ii}) the alloys Pt$_{42.5}$Cu$_{27}$Ni$_{9.5}$P$_{21}$ and Zr$_{35}$Hf$_{13}$Al$_{11}$Ag$_{8}$Ni$_{8}$Cu$_{25}$, which have very close $R_c$'s, demonstrate nonetheless quite different $\Delta\Phi_{L-X}^{max}$-values. These observations do not support the aforementioned  idea that $\Delta\Phi_{L-X}^{max}$ is related to the GFA. To our knowledge, there is currently no other data available in the literature on a possible correlation between $\Delta \Phi_{L-X}^{max}$ and $R_c$.

\begin{figure}[t]
\center{\includegraphics[scale=1.1]{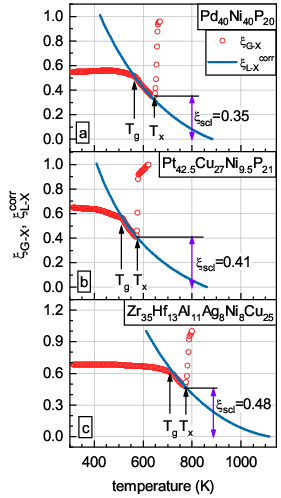}}% Here is how to import EPS art
\caption[*]{\label{Fig6.eps} Temperature dependences of the order parameter for indicated  undercooled melts ($\xi_{L-X}^{corr}$) and glasses  ($\xi_{G-X}$). The values of $\xi_{scl}$, glass transition $T_g$ and crystallization onset $T_{x}$ temperatures  are  indicated. }
\end{figure}

\subsection {Dimensionless entropy-based parameter of structural ordering and critical cooling rate }

On the other hand, it was recently found that MGs' excess entropy correlates with the glass-forming ability of corresponding melts \cite{MakarovJETPLett2024,MakarovScrMater2024}. In the view of this, it is reasonable to check if there is any correlation between the melt/glass excess entropy  and the critical cooling rate $R_c$ for the alloys under consideration.

Following the ideas presented in Ref.\cite{MakarovScrMater2024}, one can introduce dimensionless parameters of structural order $\xi_{L-X}^{corr}$ for an undercooled melt and corresponding glass $\xi_{G-X}$ as

\begin{equation}
\xi_{L-X}^{corr}(T)=1-\frac{\Delta S_{L-X}^{corr}(T)}{\Delta S_{melt}}, \label{Xi_{L-X}^{corr}}
\end{equation}
\begin{equation}
\xi_{G-X}(T)=1-\frac{\Delta S_{G-X}(T)}{\Delta S_{melt}}, \label{Xi_{G-X}}
\end{equation}
where $\Delta S_{L-X}^{corr}$ and $\Delta S_{G-X}$ are determined by Eqs (\ref{S_{L-X}^{corr}}) and (\ref{S_{G-X}}), respectively, $\Delta S_{melt}$ is the entropy increase upon heating from the solidus $T_{S}$ to liquidus $T_{L}$ temperature as defined by  Eq.(\ref{DeltaSmelt}). The order parameter varies in the range $0<\xi_{L-X}^{corr},\xi_{G-X}<1$ upon changing the structure from the fully disordered (liquid-like) state with $\Delta S_{L-X}^{corr} \rightarrow \Delta S_{melt}$ or $\Delta S_{G-X} \rightarrow \Delta S_{melt}$ and $\xi_{L-X}^{corr},\xi_{G-X}\rightarrow 0$ towards the fully ordered state (crystal-like) characterized by $\Delta S_{L-X}^{corr} \rightarrow 0$ or $\Delta S_{G-X} \rightarrow 0$ and $\xi_{L-X}^{corr},\xi_{G-X}\rightarrow 1$. 

It is interesting to calculate the order parameter for a temperature corresponding to the end of the supercooled liquid range, just before the crystallization onset at $T_c$. We denote this order parameter as $\xi_{scl}$ hereafter. It was earlier shown that  $\xi_{scl}$ correlates with the melt critical  cooling rate $R_c$. An increase of the structural order given by $\xi_{scl}$ worsens the glass forming ability (i.e. increases $R_c$) \cite{MakarovScrMater2024}. It is important to check whether a similar situation holds for the alloys studied in the present work.

\begin{figure}[t]
\center{\includegraphics[scale=1.1]{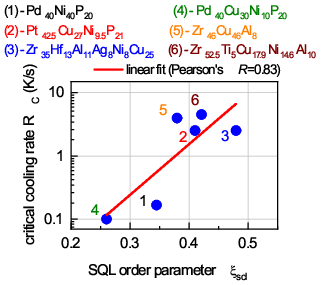}}% Here is how to import EPS art
\caption[*]{\label{Fig7.eps} Critical cooling rate $R_c$ for indicated melts as a function of the order parameter $\xi_{scl}$ calculated with Eq.(\ref{Xi_{G-X}}) for the end of the supercooled liquid region (see Fig.\ref{Fig6.eps}). }
\end{figure}

Figure \ref{Fig6.eps} shows temperature dependences of  structural order parameter for indicated undercooled melts $\xi_{L-X}^{corr}(T)$ (dark blue curves) and glasses $\xi_{G-X}(T)$ (red circles) calculated according to Eqs (\ref{Xi_{L-X}^{corr}}) and (\ref{Xi_{G-X}}), respectively. For all alloys, the situation remains nearly the same: the order parameter  $\xi_{scl}$ remain approximately constant at low temperatures, next decreases upon further heating and reaches indicated minima in the end of supercooled liquid range due to structural disordering \cite{MakarovScrMater2024,AfoninJALCOM2024b}. Continued heating of glasses leads to an increase in $\xi_{G-X}(T)$ to unity at $T=T_{cr}$ because of  crystallization. For undercooled melts, the order parameter  $\xi_{L-X}^{corr}(T)$ coincides with $\xi_{G-X}(T)$ in the  the supercooled liquid state (i.e. at $T_{g}\leq T \leq T_{x}$). At higher temperatures $T>T_x$, the order parameter of $\xi_{L-X}^{corr}$ decreases with temperature and reaches zero at the melting point, as expected.

Figure \ref{Fig7.eps} gives the critical cooling rate $R_c$ as a function of the order parameter $\xi_{scl}$ for the alloys under study. A clear increase of $R_c$ with $ \xi_{scl}$  is seen. The best glass-forming material, Pd$_{40}$Cu$_{30}$Ni$_{10}$P$_{20}$, exhibits the smallest $\xi_{scql}$.
 Similar $\xi_{scl}(R_c)$-dependence was previously found for a number of MGs \cite{MakarovScrMater2024}. Thus, structural disordering in the supercooled liquid state of glass reflects the ease of melt glass formation.  The smaller the order parameter $\xi_{scl}$ (i.e. the closer the structure to that of the equilibrium liquid), the lower the $R_c$ is.   

\section {Conclusions}

The study reports the first comparative analysis of  temperature evolution of the excess thermodynamic potentials (enthalpy $\Delta H$, entropy $\Delta S$ and Gibbs free energy $\Delta\Phi$) of three alloys taken as an example, Pd$_{40}$Ni$_{40}$P$_{20}$, Pt$_{42.5}$Cu$_{27}$Ni$_{9.5}$P$_{21}$ and Zr$_{35}$Hf$_{13}$Al$_{11}$Ag$_{8}$Ni$_{8}$Cu$_{25}$, in the undercooled molten state and solid glassy state. Excess thermodynamic potentials of undercooled melts with respect to the crystalline state, $\Delta H_{L-X}$, $\Delta S_{L-X}$ and $\Delta \Phi_{L-X}$, are taken from the literature. For the glassy state, the excess enthalpy $\Delta H_{G-X}$, entropy $\Delta S_{G-X}$ and Gibbs free energy  $\Delta \Phi_{L-X}$ were calculated on the basis of calorimetry data using the method proposed recently.       

It is found that temperature dependences of the excess enthalpies for undercooled melts and glasses, $\Delta H_{L-X}(T)$ and $\Delta H_{G-X}(T)$, coincide in the supercooled liquid range, i.e. at temperatures $T_{g}\leq T \leq T_{x}$ (where $T_g$ and $T_x$ are the glass transition and crystallization onset temperatures, respectively). Temperature dependences of the excess entropies, $\Delta S_{L-X}(T)$ and $\Delta S_{G-X}(T)$ also coincide in this temperature range but only after important corrections related to the calculations of the melting entropy. The Gibbs free energies of undercooled melts $\Delta \Phi_{L-X}(T)$ are also equal to glass Gibbs free energies $\Delta \Phi_{G-X}(T)$ in the supercooled liquid range provided that the same entropy corrections are applied.

We also studied the effect of the excess Gibbs free energy and entropy on the glass-forming ability (GFA), which is directly characterized by the melt critical cooling rate $R_c$. First, we found no clear relationship between $R_c$ and melt excess Gibbs free energy at the maximum of the  $\Delta \Phi_{L-X}(T)$-dependence, contrary to what is suggested in the literature.

On the other hand, on the basis entropy data, we introduced  a dimensionless order parameter $\xi$, which changes in the range $0<\xi<1$. It is shown that temperature dependences of this parameter determined with undercooled melt and glass entropy data coincide in the supercooled liquid range. It is also found that the order parameter $\xi_{scl}$ calculated for the end of this temperature range (i.e. for a temperature just below $T_x$) correlates with the critical cooling rate: a decrease of structural order given by  $\xi_{scl}$ results in a decrease of $R_c$ (i.e. leads to better GFA), in line with literature data.  

\section*{Declaration of competing interests}

The authors declare that they have no known competing financial
interests or personal relationships that could have appeared to influence
the work reported in this paper.

%\textbf{Data availability}

%Data will be made available on request.

\section*{Acknowledgements}
The work was supported by the Russian Science Foundation under the project No. 23-12-00162.

\end{document}